# The Effectiveness of Embedded Values Analysis Modules in Computer Science Education: An Empirical Study

Matthew Kopec[*], Meica Magnani, Vance Ricks, Roben Torosyan, John Basl, Nicholas Miklaucic, Felix Muzny, Ronald Sandler, Christo Wilson, Adam Wisniewski-Jensen, Cora Lundgren, Ryan Baylon, Kevin Mills, Mark Wells

## Abstract

Embedding ethics modules within computer science courses has become a popular response to the growing recognition that CS programs need to better equip their students to navigate the ethical dimensions of computing technologies like AI, machine learning, and big data analytics. However, the popularity of this approach has outpaced the evidence of its positive outcomes. To help close that gap, this empirical study reports positive results from Northeastern University's program that embeds values analysis modules into CS courses. The resulting data suggest that such modules have a positive effect on students' moral attitudes and that students leave the modules believing they are more prepared to navigate the ethical dimensions they will likely face in their eventual careers. Importantly, these gains were accomplished at an institution without a philosophy doctoral program, suggesting this strategy can be effectively employed by a wider range of institutions than many have thought.

## Acknowledgements

We would like to thank Katherine Simeon from Northeastern's Center for the Advancement of Teaching and Learning through Research, who provided initial assistance with the survey questions. A generous grant from the Mozilla Foundation's Responsible CS Challenge made this study possible.

---

[*] Corresponding Author: matthewckopec@gmail.com or m.kopec@northeastern.edu

# 1. Introduction

There is a growing consensus that computer science students ought to be better prepared to navigate the ethical dimensions of computing technologies, in particular those raised by artificial intelligence, machine learning, and big data (Borenstein and Howard, 2021; Burton et al., 2017; Goldsmith and Burton, 2017). Nonetheless, university degree programs have thus far struggled to provide that training (Borenstein and Howard, 2021; Spradling et al., 2008), due to a number of mutually reinforcing roadblocks (Tugend, 2018). The Computer Science ('CS') curriculum at most universities is notoriously packed, making for very limited space to include full courses on computing technology ethics (Singer, 2019). Because the CS curriculum often doesn't require a full course in CS ethics , programs consequently lack specialized faculty to teach such courses if they are eventually added (Keefer et al., 2014). CS faculty without a specialization in ethics may not be properly trained or experienced enough to teach ethics courses on their own (Johnson, 1994), and, even if they are, many would rather not (Quinn, 2006). This lack of instructor capacity or interest likely makes its way into the classroom, since students often find required tech ethics courses annoying, unnecessary, or generally irrelevant to their CS training (Skirpan et al., 2018). And, as often suggested by proponents of the 'ethics across the curriculum movement' (Mitcham and Englehardt, 2019; Weil, Vivian, 2003), one-off ethics courses that are completely disconnected from the overall CS curriculum are unlikely to have the lasting impacts necessary to realize positive outcomes across the field (May and Luth, 2013; Skirpan et al., 2018). Therefore, the challenge is to find an educational program that will fit within an already crowded curriculum, that will be taught by competent and motivated instructors, that will quickly exhibit its value to students, and that will make repeated contact with students throughout their course of study.

One increasingly popular strategy to meet the above challenge is to embed short ethics modules, typically consisting of one or two relatively self-contained class sessions taught by either philosophers trained in ethics or by interdisciplinary teams, into courses across the CS curriculum (Bezuidenhout and Ratti, 2021; Furey and Martin, 2019; Grosz et al., 2019; Horton et al., 2022). Since ethics modules don't replace any existing courses, it is easier to find space for them in the curriculum. CS faculty who lack the desire or the expertise to teach the ethics modules can lean on the expertise of ethics faculty, postdocs, or graduate students from other disciplines like philosophy, either to run the modules entirely or to do so collaboratively with the CS faculty member. This, in turn, places an educator who is well-versed in ethics and excited about the ethics material in front of the students, which should have positive effects on the level of student enjoyment, engagement, and overall appreciation for the ethical considerations raised by computing technologies. Finally, when such modules are contextualized for a range of courses that span the CS curriculum, CS students will receive repeated exposure, reinforcing any ethical growth the modules generate individually. The theoretical basis for this approach is thus very compelling, which goes a long way to explain why prominent universities such as Harvard, MIT, Stanford, and Toronto (among others) are rushing to implement or expand such programs.

The above story is compelling but merely theoretical at this point. Although some evidence for the effectiveness of embedded ethics curricula has been collected (Coldwell et al., 2020; Horton et al., 2022), overall, the popularity of these programs has greatly surpassed careful assessment of the strategy's effectiveness. The present study aims toward closing this gap in the evidence.

Since 2019, instructors from Northeastern's Department of Philosophy, Khoury College of Computer Science, and Ethics Institute have been running ethics modules across a range of Computer Science, Data Science, and Cybersecurity courses (initially supported through Mozilla's Responsible Computer Science Challenge). These modules focus on teaching "Value Analysis in Design" as a way of fostering a set of reflective attitudes and skills that we believe are a key component in both "Value Sensitive Design" (or VSD) and ethically reflective technology design and development more generally. During the fall semester of 2021, we ran a pre-registered empirical study to examine the effectiveness of these modules. First, the study was designed to test the effect of the modules on certain normative attitudes that we see as central to promoting the ethical design, development, and deployment of computing technologies. Second, the study examined whether these modules increased the students' perceived interest in, and perceived skill at, navigating the ethical considerations raised by their expected field of work.

What we found was that these short modules, which lasted no more than two class periods, were enough to have a positive impact on those attitudes, to generate a greater motivation to address ethical considerations in their field, and to foster a greater expected self-mastery to do so. Importantly, Northeastern's team was able to accomplish these results without the resources associated with a doctoral program in philosophy, which suggests that the strategy may be effectively employed at a wide range of institutions.

This study is merely an early step toward showing that the embedded modular approach to CS ethics education is successful. In Section 6.2, we discuss some necessary future work, including paring survey studies like this with aspects of diagnostic tools developed for other purposes – e.g., the MJI, DIT, TESSE, EERI, etc. – which resource restrictions made impractical for the present study.

## 2. Values Analysis in Design

It is now well established that AI, machine learning, big data, and other emerging computer technologies raise significant ethical concerns and, at the same time, provide great promise for enhancing human flourishing (Barocas and Selbst, 2016; Crawford and Calo, 2016; Mittelstadt et al., 2016; Whittaker et al., 2018).

Unfortunately, there remains a great deal of disagreement over what the best methods to mitigate those concerns or foster that flourishing might be and what the proper role of the researchers, designers, and developers of such tools should be in the process. But one point of widespread agreement is that making those who are responsible for building these technologies more cognizant of, and responsive to, the ethical considerations that these tools may raise is a step in the right direction. This preliminary step is the core goal of Northeastern's modular ethics education program for computer science students.

There are a number of roadblocks to building this kind of ethically responsible mindset, including social, psychological, motivational, economic, and attitudinal factors, and a short educational program obviously can't effectively tackle all of them. But our contention is that there are a number of core attitudes that impede our students' progress toward becoming ethically responsible tech designers. In our team's collective classroom experience, many students tend to believe that the moral concerns raised by new technologies arise only at certain points in the adoption and implementation process, and that they generally don't arise until well after the designer's work is done. For example, our students often believe that new technologies are simply tools, and any ethical concerns that such tools raise are therefore associated with the user, not the designer or builder. (Admittedly, it is not just students who think this; see (Miller, 2021; Pitt, 2014) for discussion.) Many of those students who are willing to admit that design choices raise moral concerns still believe that those concerns should be dealt with downstream, e.g., by compliance or oversight boards, which would similarly imply that the designers need not bother with any of the moral concerns themselves. Many of the students willing to admit that moral concerns do arise during the design process nonetheless believe that the ethical solutions are straightforward, since one can simply check what the relevant legal statutes dictate to determine what one ought to do. Even those who do not equate what is moral with what is legally permissible often think there is very little on the ethical side to puzzle through, due to their conviction that anything of moral concern with a new technology will either have a purely technical solution, that further technological advancement will solve any earlier problems caused, or that there is some concrete calculus that provides a clear-cut way to adjudicate between supposed ethical tradeoffs.

One core objective of Northeastern's Values Analysis in Design (VAD) modules is to counter each of these problematic attitudes. First, VAD modules emphasize that computing technologies are not merely value neutral tools, but, rather, that values are enmeshed across all stages of the process of technology design, development, and implementation (Nissenbaum, 1998; van de Poel, 2021; Verbeek, 2006). A particular emphasis is placed on how design occurs within a value-laden, socio-technical context (Feng and Feenberg, 2008; van de Poel and Royakkers, 2011), how design choices can have serious socio-political effects (Friedman and Nissenbaum, 1996), and how artifacts can, themselves, "have politics" (Winner, 1980). Second, VAD modules emphasize the importance of weighing the moral implications of each choice made in the design process (Friedman and Hendry, 2019; van de Poel and Royakkers, 2011), which runs counter to the common attitude that ethics ought to be dealt with downstream by legal or compliance teams. Third, VAD modules attempt to dispel the common misconception that all ethical matters are settled by the law (Hart, 2012; Raz, 1985). Finally, the VAD modules emphasize that the ethical concerns raised by computing technologies are not merely technical problems with solely technical solutions (Mittelstadt, 2019), showing, instead, that tackling them requires a diverse range of expertise offered from an array of disciplinary perspectives (Friedman and Hendry, 2019).

VAD modules are a crucial component of Northeastern's educational program in computing ethics, which, as a whole, aims to teach CS students about Value Sensitive Design (often referred to as "VSD"). Value Sensitive Design is "An approach that aims at integrating values of ethical importance in a systematic way in engineering design"(van de Poel and Royakkers, 2011), and it is "considered by many as the most comprehensive approach to account for human values in technology design" (Winkler and Spiekermann, 2021). Value Sensitive Design has been described fully in the literature, so we will not repeat all the details here (see Winkler and Spiekermann, 2021). But to situate our approach with that literature: we see VAD modules as a core component of the "conceptual" investigation portion of the tripartite methodology that is common among Value Sensitive Design approaches (Friedman et al., 2013; Friedman and Hendry, 2019; Winkler and Spiekermann, 2021). Although VAD modules might briefly touch on some aspect of empirical and technical investigations – the other members in the tripartite methodology – the primary focus is on cultivating a distinctively normative skillset for navigating the ethical considerations technologies raise. Put another way, VAD modules primarily tackle the normative or philosophical questions that arise in Value Sensitive Design, as opposed to attempting to subsume the whole methodology into a single intervention.

Our contention is that philosophers are especially (though not uniquely) well-suited to help students tease out the conflicting moral, social, and political values that arise in the technological design process. Philosophers are also well suited (though, again, not uniquely) to guide students through the critical evaluation of technology's impact on society, which is another key aspect of Value Sensitive Design. In accord with the "ethics across the curriculum" approach (Cruz and Frey, 2003; Mitcham and Englehardt, 2019; Weil, Vivian, 2003), we believe that students will be better able to employ Value Sensitive Design if they have repeated exposure to well-trained instructors who guide them through these conflicts and critical evaluations. Thus, Northeastern's program aims to divide the labor of a full Value Sensitive Design approach, with trained ethicists leading instruction of the portions they are most skilled at leading.

For those who are sympathetic to the prominent critiques of VSD (Davis and Nathan, 2015; Le Dantec et al., 2009; Manders-Huits, 2011; Yetim, 2011), it is important to note that VAD modules, as a curricular intervention, are not essentially tied to the VSD framework. Rather the core aspects of VAD modules can feed into a wide variety of different values-oriented approaches to CS design. These include Adversarial Design (Disalvo, 2015), AI for Social Good (Floridi et al., 2018, 2021), Critical Technical Practice (Agre, 1997), Design for Values (Aizenberg and van den Hoven, 2020; van den Hoven et al., 2015), Ethics by Design (Gerdes, 2021), Ethics Parallel Research (Jongsma and Bredenoord, 2020), Ideologically-Embedded Design (Détienne et al., 2019), Morality in Design (Verbeek, 2006, 2008), Reflective Design (Sengers et al., 2005), Responsible Research and Innovation (Stilgoe et al., 2013), Values at Play (Flanagan et al., 2005), Value-Centered Design (Cockton, 2005), Values-Led Participatory Design (Iversen et al., 2012; van der Velden and Mörtberg, 2015), and the examination of Values Levers (Shilton, 2013, 2018). Essentially, any approach that requires students to deeply and critically examine the individual, social, and moral values involved in the design process would benefit from the core aspects of VAD modules. The only approaches to ethical computing that would clash with these modules are those based on one of the problematic assumptions discussed earlier in this article. So, if an approach assumes that values do not enter at all stages in the design process, that decisions of ethical importance arise only after the technology has been released into the world, that ethical design is simply determined by the law or compliance officers, or that ethical design is a purely technical matter and expertise from other fields is thus irrelevant, then VAD modules may not be compatible with such an approach. Since we contend that such approaches are all deeply flawed, we don't see this as a problem for our program.

## 3. VAD Modules in Practice

Since its inception in 2019, the core VAD instructional team has changed over time and has at various points included three postdoctoral researchers, one staff research administrator, one computer science faculty member, and four faculty members in philosophy. In addition to the core team, the computer science instructors leading the courses into which modules were embedded have often played an important role in structuring and running the modules, although the level of such involvement has been highly variable. Starting in the fall of 2021, the semester during which we ran this study, development of content modules and core instruction has been led by Professors Meica Magnani and Vance Ricks, both of whom are philosophers with doctorates, who joined the program through newly created permanent full-time teaching faculty positions that are jointly appointed between Philosophy and Computer Science. Although Ricks had prior formal CS training and previously co-taught a course in programming, Magnani's prior engagement with CS was solely on the ethics side, through a prior position with Harvard's Embedded EthiCS program. Developing and running the modules accounts for only a portion of these faculty's teaching loads each semester – the equivalent of one course. Thus, the model for administering Northeastern's program is substantially different than the other better-known programs mentioned earlier.

For fall 2021, Magnani and Ricks ran VAD modules in a total of seven sections of five distinct courses: CS4100 *Foundations of Artificial Intelligence*, CS4120 *Natural Language Processing*, CS4550 *Web Development*, CY2550 *Foundations of Cybersecurity*, and DS4400 *Data Mining/Machine Learning 1*. All of these courses ran at the undergraduate level, although two (CS4120 and CS4550) were run concurrently with a graduate version. All class sessions were one hour and forty minutes long and all sessions had a remote attendance option. CS4550 was the only course without an in-person attendance option. CY2550 was run as two non-concurrent sections by the same cybersecurity instructor, while two different sections of CS4100 were run by two different main CS instructors. CY2550 was also unique in that the VAD module consisted of only one class session, whereas all others had two. Although this was the first semester that Magnani and Ricks ran these modules, piloted versions of the modules were previously run by other team members for each course. The total starting enrollment for these courses was 462.

Below, we sketch the details of two of these modules to give the reader some idea of how a VAD module might run, some of the topics that might be covered, etc.

3.1. CS4100 Foundations of Artificial Intelligence

The course CS4100 *Foundations of Artificial Intelligence* introduces students to the fundamental technical problems, theories, and algorithms in the field of artificial intelligence. The VAD module embedded into this course ran over two consecutive class sessions for one section but, for the other section, they ran a month apart. Magnani ran all module sessions, which all lasted an hour and forty minutes total. The module focused on introducing students to the moral and social considerations raised by involving AI in our judgment and decision-making.

In the first session, students were introduced to the Moral Machine project (https://www.moralmachine.net/), a data collection platform run out of MIT, which was designed to gather ethical intuitions about how self-driving cars should handle ethical trade-offs - e.g., stay straight, and kill two people jaywalking, or swerve, and kill one person crossing legally (Awad et al., 2018). First in small groups, and then as a whole class, students considered and discussed a list of ethical trade-offs presented to them on the platform, tasked with identifying what they took to be ethically salient factors in each decision. The concepts of normative versus descriptive ethics were introduced to frame a whole class discussion about what factors should inform our models of ethical reasoning and whether the Moral Machine (as a survey of public intuition) is a good guide for designing how self-driving cars should operate. Students were then given a brief introduction to some of the foundational components of Value Sensitive Design, and how this might serve as a different approach to programming self-driving cars. A short lecture then gave students guidance on how to perform a stakeholder analysis, including how to identify the various stakeholders, what interests might be at stake for them, and how these interests might come into conflict. The class was then asked to suggest some technical or policy interventions that might mitigate such conflict of interests, and the whole class critically evaluated those interventions through open discussion. The session concluded with an open discussion about how the widespread implementation of self-driving cars might impact current practices, institutions, norms, and values. The total activity type breakdown was roughly 30 minutes of lecture, 30 minutes of small group tasks, and 40 minutes of full class discussion.

The second session in this module examined how ethically problematic biases can enter into the development and deployment of automated decision-making tools. First, students were given some background on the sources of bias in algorithms, including biases that stem from institutional goals, individual designer bias, and biased training data that are tainted by past social injustices. Students were then given a small group hands-on activity where they worked through a number of automated decision-making examples, e.g., facial recognition tools and the Allegheny Family Screening Tool (AFST), to diagnose the possible sources of bias in these models, anticipate resulting harms, and identify socio-historical features that might amplify any biases they may contain. The session closed with the whole class discussing what the small groups had uncovered while working through the task. The total activity type breakdown was roughly 15 minutes of lecture, 45 minutes of small group tasks, and 40 minutes of full class discussion.

At the very end of the second session, students were given an individual out-of-class assignment where they would choose one AI application to perform a values analysis on, e.g. autonomous weapons systems, health care or domestic service robots, or a choice of their own, in order to solidify what they had learned over the two sessions. The values analysis included such tasks as giving examples of direct and indirect stakeholders, possible value conflicts between these stakeholders, possible worst-case scenarios from widespread deployment of the application, etc. That assignment was expected to take roughly an hour, and it was to be completed as part of their regular problem set for the following week. Graduate TAs from CS graded the assignments based upon a rubric designed by Magnani.

### 3.2. DS4400 Data Mining/Machine Learning 1

The course DS4400 *Data Mining/Machine Learning 1* introduces students to supervised and unsupervised predictive modeling, data mining, and machine-learning concepts, using tools and libraries to analyze data sets, build predictive models, and evaluate those models' fit. The VAD module embedded into this course ran over two consecutive one hour and forty minute class sessions, both run by Ricks. The module focused on introducing students to different philosophical accounts of fairness, examined how those accounts might differ from formal criteria of fairness used in CS, and examined the case for rejecting some "perfectly" predictive software models for ethical or sociopolitical reasons.

The first session began with an introduction to some recent controversies involving machine learning tools, e.g., Zoom's automated background failures for dark skinned individuals (Costley, 2020) and the hiring algorithms Amazon developed that showed biases against women job seekers (Dastin, 2018). The students were then broken up into small groups and asked to devise some determinants of fair treatment and fair outcomes and discussed the results as a whole class, as a way of introducing the difference between "patterned" accounts and "procedural" accounts of fairness. Students returned to small groups to play the "Survival of the Best Fit" (SOTBF) game, which is a web-based game designed by a group of students from NYU Abu Dhabi, where players attempt to design a fair hiring procedure (available at https://www.survivalofthebestfit.com/). The results were then discussed as a whole class to determine whether any of the groups were able to devise a hiring procedure that met both accounts of fairness. This was followed by a lecture covering different possible sources of unfairness (e.g., based on purpose, data collection, or distribution of burdens) and possible ways to mitigate it, including various concrete examples of biased algorithmic systems (e.g., the COMPAS recidivism prediction software (Chouldechova, 2017)), interspersed with whole class discussion breaks. The total activity type breakdown was roughly 40 minutes of lecture, 40 minutes of small group tasks, and 20 minutes of full class discussion.

The second session began with a quick review of the previous class session along with a short whole class discussion about the difference between legal norms of fairness and moral norms of fairness. Students were then introduced to some of the core aspects of Value Sensitive Design, including how to determine who the direct and indirect stakeholders might be, which moral and social values might be at stake in the development and deployment of certain technologies, how to identify and adjudicate possible value tradeoffs, and the tripartite division of disciplinary labor (between conceptual, empirical, and technical investigations). This was followed by a whole class discussion of how to perform a values analysis on the SOTBF game in addition to a hypothetical university admission algorithm, followed by a general discussion of how society might best avoid coming to rely upon unfair algorithms. The total activity type breakdown was roughly 1 hour of lecture, 10 minutes of small group tasks, and 30 minutes of full class discussion.

The students were also given an at-home assignment, to be completed in groups, where they performed a values analysis on the university admission algorithm discussed above. Graduate TAs from CS graded the assignments based upon a rubric designed by Ricks.

## 4. Study Methodology

In the present pre-registered study (https://osf.io/vtsmp), we examined the success of VAD modules at promoting a proper subset of the overall learning outcomes, namely, their impact on certain ethically relevant attitudes. As alluded to earlier, VAD modules aim to foster four attitudes needed for more ethically responsible technology design. These four attitudes may be stated as follows:

> 1. Value laden choices appear throughout the technology design process.
> 2. The designer of a technology ought to consider the moral implications of each choice they make.
> 3. The moral acceptability of a technology design choice is not simply settled by the relevant legal statutes.
> 4. Since the moral concerns raised by technology design don't always have purely technical solutions, a range of disciplinary expertise is needed to properly address them.

Established, tractable methods for assessing the effects of interventions on attitudes have become commonplace in educational psychology (Lamprianou and Athanasou, 2009) and in the behavioral sciences more broadly (Fishbein and Ajzen, 1972). The present study applied some of those methods to examine whether the modules have the desired effects on those four attitudes.

To assess the effect of module completion on those attitudes, we settled upon a one-group pre-intervention / post-intervention survey methodology (Price et al., 2017). Respondents were prompted to "Please select how much you agree or disagree with each of the following statements," using a 5-point Likert scale, with anchors labeled "Strongly Disagree," "Somewhat Disagree," "Neither Agree nor Disagree," "Somewhat Agree," and Strongly Agree"(Drinkwater, 1965). During initial survey design, we randomized whether agreement or disagreement with the relevant claims was the desired outcome. This procedure determined that agreement was the desired outcome for 1, 2, and 4 and disagreement the desired outcome for 3. Instead of asking the students' agreement with the claims above directly, we chose close relatives of the claims - statements that one would expect someone to agree or disagree with to the same extent as someone agrees or disagrees with the originals. This was done to avoid the possibility of priming effects (Reis and Judd, 2014), in case those claims above were explicitly asserted verbatim during the session.

The result was a pre-post module survey of levels of agreement "with each of the following statements":

> A1. All instances of technology design are value laden at multiple points in the design process.
> A2. When making technological design choices, it is important to consider the moral implications of each choice.
> A3. All technology design choices that are fully permitted according to the law are morally permissible.

> A4. In order to properly address the moral concerns raised by technology, designers need to engage with the work of experts from a range of disciplines.

To determine the presence of other confounding survey effects, the following statement was added to this bank of questions:

> R1: Before enrolling at Northeastern, I was confident that I only wanted to major in Computer Science or a closely related field.

We expected no movement for this question. Therefore, any pattern of movement on responses to it could indicate reduced robustness of our results, e.g., suggesting a notable amount of noise in the data or a right-hand bias in the survey methodology (Reis and Judd, 2014).

We also examined what impact the students themselves believed the modules had on their long-term motivation and ability to tackle the ethical considerations raised by the computing technologies covered in the course. Such use of self-report is an established measure in educational psychology and pedagogical research (Fredricks and McColskey, 2012). Furthermore, there is support in the pedagogical literature for mixing such post-intervention-only values along with change values when constructing an overall analysis of an intervention's impact on learning outcomes (Higgins et al., 2019). Based on these methodologies, we included five additional self-assessment questions as part of the post-module surveys, using the same 5-point Likert scale for disagreement to agreement as above. The first four questions targeted self-reported gains we hoped students would agree they had made:

> A5: The Value Sensitive Design module has improved my ability to notice the possible ethical implications of the material covered in this course.
> A6: The Value Sensitive Design module has given me a greater appreciation for the broad range of ethical concerns that data and computing technologies can raise.
> A7: The Value Sensitive Design module has motivated me to help address any ethical concerns I encounter in my career.
> A8: After taking the Value Sensitive Design module, I expect to be better able to track down the assistance necessary to navigate any ethical concerns I encounter in my career.

As a check for experimenter bias, we included an additional question that we expected students to be no more than neutral on if answering sincerely (Reis and Judd, 2014):

> R2: The Value Sensitive Design module has improved my computer programming skills.

Given these survey questions, we essentially had 9 initial hypotheses that guided the experiment. For the first five, we hypothesized that students would agree more with A1, A2, and A4 on their post-module survey than on their pre-module survey, agree less with A3 on their post-module survey, and that the movement on each of these would be greater than any movement on R1. (These are hypotheses H1-H5 on the preregistration.) For the next four, we hypothesized that students would agree more with each of A5-A8 than they did with R2 on their post-module survey. (These are H6-H9 on the preregistration. H1-H9 were the only pre-registered hypotheses.)

To test the hypotheses above, we constructed two versions of the survey, one to be completed before students attended their first module (which included A1-A4 and R1), and one to be completed after (which included A1-A8, R1 and R2). All of those responses were compulsory. Both surveys included a bank of demographic questions, including questions about prior coursework in philosophy, and the post-module survey included some questions to assess attendance (for purposes of exclusion) and the amount of total out-of-class time spent on the module (to assess effort). For all modules, successful completion of the exit survey was required to receive the full portion of course credit assigned to the module, which in some cases was tied to further completion of assignments to be completed after attending the modules. Since each course had an online attendance option, surveys were delivered to students electronically to be taken independently (making in-person and virtual attendees indistinguishable in the data).

Because some students were enrolled in multiple courses running modules, all students were instructed to complete the pre- and post-module surveys attached to the course for which the module started earliest in the semester. The survey links were delivered to students by the CS instructor's preferred route, which varied from course to course (e.g., email, Piazza, Canvas, etc.). Those multiply-enrolled students were sent a follow-up email to further clarify the requirement. Pre-module surveys attached to a module were delivered to students the week before the module and scheduled to close 15 minutes after the start of the first module session for that course (to allow module instructors the opportunity to request completion at the very start of the session). Post-module surveys used the same instructor-chosen distribution method, were delivered to students either after completion of the module or any associated assignments (depending on the course), and were set to close within two weeks of the final module session for that course. All surveys were run through a secure Northeastern University licensed Qualtrics account, with different surveys being constructed for each course, for tracking and exclusion purposes. Students were informed that simply following the relevant links and then refusing to consent to participating in the study would count as "successful completion" of the survey. So, participating in the study was entirely voluntary.

Prior to distribution of surveys, Northeastern's IRB was fully informed about the details of the study, but determined that since it was an educational improvement project, the study did not constitute Human Subjects Research according to the relevant statutes and therefore did not warrant full IRB review. After that determination, the full study, including hypotheses to be tested, data exclusion criteria, and initial analyses to be run, was pre-registered on OSF. The exclusion criteria were designed to exclude anyone not enrolled in the specified course and any enrolled undergraduate students who completed surveys too quickly (30 seconds for pre- and 90 seconds for post-), who didn't attend any module sessions, who were possibly exposed to a module prior to submitting their first pre-module survey due to multiple enrollments, who submitted survey responses for multiple courses, or who failed to submit exactly one pre- and one post-module survey for the same course. If any exclusion criterion was triggered for a student on any survey, all of that student's data was removed from the study. A power analysis was run for informational purposes only, since the pre-exclusion sample size was capped by the total enrollment of the courses.

Section 3 contains some additional details relevant to study design – e.g., differences between the various courses, available attendance options, whether modules were piloted, etc. And for full details about all survey questions, pre-registration documentation, IRB documentation, exclusion criteria and the Python code to execute them, etc., see Supplementary Materials.

## 5. Results

Of the total unique student enrollment of 431 in all seven sections of the five courses that ran modules, we received pre-module responses from 329 unique enrolled students and post-module responses from 335 unique enrolled students, for response rates of 76% and 78% respectively, substantially above current standards in the social sciences (Holtom et al., 2022). After processing all exclusion criteria, 189 paired pre- and post-module responses remained, or roughly 44% of total enrollment. (For a specific breakdown of these exclusions, see Supplementary Materials.) Such strict exclusion criteria were necessary to promote a high-quality dataset of sincere responses, and all exclusion criteria were listed in the pre-registration for the study.

To reiterate, we hypothesized that student responses would increase on questions A1, A2 and A4, and decrease on A3. The post-exclusion data were consistent with all of these hypotheses, and confirmatory for three of them. The pre-post mean value for A1 increased from 3.87 to 4.21 ($p < 0.001$, $r_{bc} = 0.43$), for A2 it increased from 4.57 to 4.65 ($p < 0.05$, $r_{bc} = 0.33$), and for A3 it decreased from 1.80 to 1.61 ($p < 0.01$, $r_{bc} = -0.39$). (All reported p values are Wilcoxon rank sum; Student T test values were similar. Reported effect sizes are the matched-pair rank-biserial correlation values for each.) The mean value for A4 increased only a modest amount, from 4.54 to 4.61 ($p = 0.1$, $r_{bc} = 0.17$). Although we expected no change in responses for R1, since the question asks students about how much they agreed with a claim about their confidence in majoring in CS or a related field at a prior date, we unfortunately did see movement on that question as well (increasing from 3.63 to 3.71). These values are all represented in Figure 1. Even when comparing movement in the target attitudes against relative movement in R1, these results still pass significance for two of the four target questions (A1 with $p < 0.01$ and A3 with $p < 0.05$). Finally, a regression was conducted to assess whether there were any notable, possibly confounding relationships between changes in responses and student demographics or educational background, and no significant correlations were found. (See Supplementary Materials for details.)

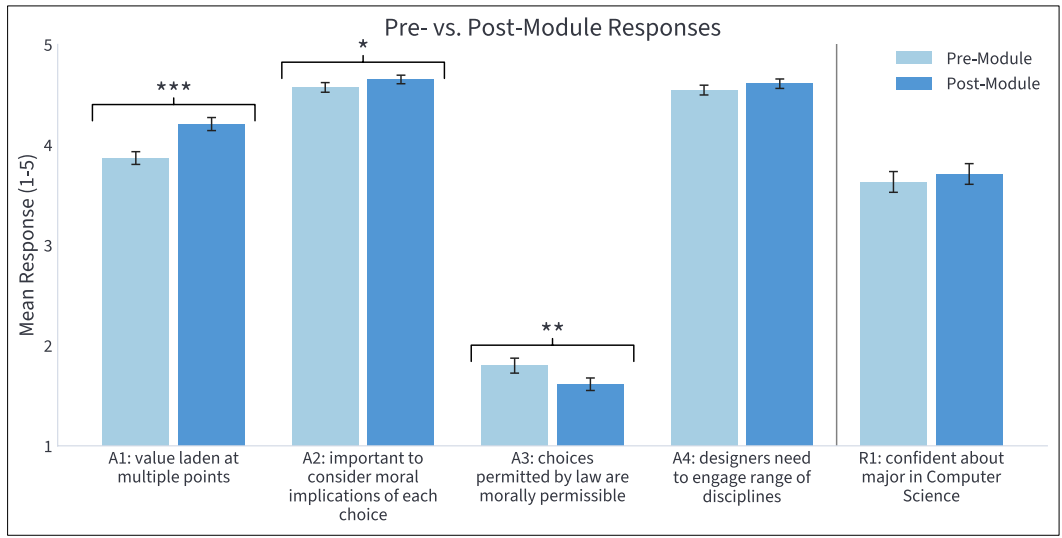

Figure 1: Responses for target questions A1-A4 and R1 that were asked on both pre- and post-module surveys. Question texts are paraphrased, see body text for exact wording. Error bars are 1 SE, and n = 189 for all questions. (*** $p < 0.001$, ** $p < 0.01$, * $p < 0.05$)

Having any statistically significant impact on student attitudes from such a short educational program is notable. In addition, there were a number of features of the study that likely weakened the results. First among these was the very high number of exclusions. This, we assume, traces back to the fact that there were minimal incentives for proper completion, leaving many students without the motivation to ensure they were taking the correct survey for the correct course the correct number of times. Second, the 5-point Likert scale may have been too blunt an instrument, due to the high number of extreme responses on the pre-module surveys. The concern here is that if a high proportion of pre-module responses on these four attitudes are already as high and in the hoped-for direction as they could possibly go, then there simply is no room to move for many of the students. The data suggest this probably affected our results: on A1, where we received our strongest result, only 24.9% held the most extreme target value, compared with 64% for A2, 50.3% for A3, and 61.4% for A4. It is, of course, a welcome sign that many of our students already seem to have held these important attitudes prior to taking the modules, but this substantial group of students nonetheless make it harder to assess the overall effect of the modules. If we had used an instrument with more sensitivity, e.g., a 7-point Likert scale, we might have registered more of an effect.

For the post-module-only questions (A5-A8), the results were much more striking. For each question, the mean value was well above the midpoint value of 3 (A5: 4.13, A6: 4.23, A7: 4.03, A8: 3.88; $p < 0.001$ for each; $r_{bc} > 0.95$ for each). Question R2 was included to assess whether there was any "experimenter bias", where the students are simply responding with the rosy picture they assume the survey designer wanted to see. If students tended to agree that the modules had made them better programmers, that would have been evidence that the students were attempting to give the experimenters what the latter wanted to hear. We found the contrary, with mean values on R2 of 2.58, well below the midpoint value, which suggests experimenter bias isn't a legitimate concern. These values are represented in Figure 2. (We note that comparing the values on A5-A8 against R2, as suggested in the pre-registration for this study, would only strengthen our results.) These results suggest that students are indeed leaving the modules feeling as though they have better appreciation of the moral concerns raised by computing technologies, more motivation to help address those concerns, and with a better grasp of the available resources needed to do so. Although these suggestive results would need to be confirmed through longitudinal studies, they at least provide some evidence that the modules could have long-term positive effects.

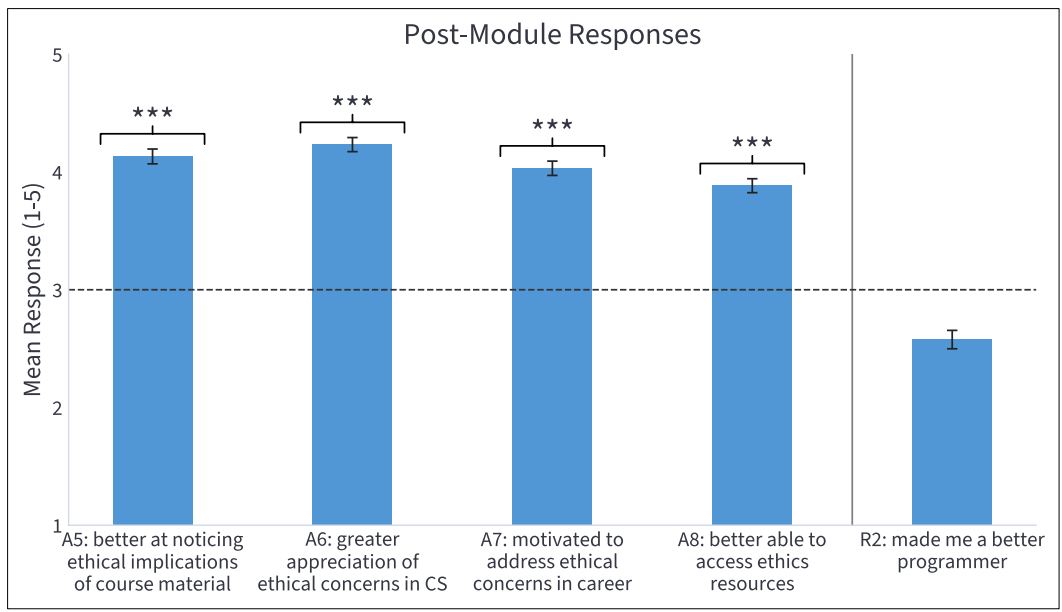

Figure 2: Responses for target questions A5-A8 and R2 that were asked only on post-module Surveys. Question texts are paraphrased, see body text for exact wording. Error bars are 1 SE, and n = 189 for all questions. (*** $p < 0.001$)

# 6. Discussion

In this study, we found that students taking our VAD modules had notable changes in some important normative attitudes and self-reported other ethically relevant effects. When measuring the four pre- versus post-module questions just in terms of movement, there was statistically significant movement on three out of four questions, and all questions moved in the direction we hoped for. It is noteworthy that we saw statistically significant movement even while using a relatively blunt instrument (the 5-point Likert scale) and excluding a large number of survey responses that triggered at least one exclusion criterion (which eliminated 146 students from the data). Although student self-reports about the positive impacts of the modules are clearly fallible, the fact that students overall seemed to believe the modules will help them notice, care about, and navigate ethical dilemmas (i.e., questions A5-8) is at least some evidence that they are effective on that front as well. Further studies will need to be done to fill this picture out, but this is solid initial evidence that embedded VAD modules can have a number of positive effects, at least in certain contexts like ours.

But as is typical in pedagogical research, the present study was not able to fully assess whether our VAD modules effectively promoted all of our learning goals – given the available resources, technological tools, and timeline – and was subject to various other limitations. We discuss these limitations below.

## 6.1. Limitations

Some aspects of the survey instrument itself introduced limitations. First, for each of the attitudes we were examining, we only used one question in the assessment. The results would have been more robust if we had used multiple questions to assess each attitude from slightly different angles. In more recent version of the survey, we have added some questions used by (Horton et al., 2022) to help with this. Second, we could have done a full validation study, including a range of comprehension questions, to ensure that all the questions were likely to be parsed correctly, which would have strengthened the weight of the results. Finally, as we discuss in more detail below, this one-off study yields limited insight into whether the movement on these attitudes will resist the kinds of social, psychological, motivational, economic, and attitudinal pressures that students will experience once they enter the workforce. A much more extensive longitudinal study, one that tracks students after they enter the workforce, would be necessary before making any fully concrete judgements about the positive societal effects of the program.

Additionally, due to some important programmatic differences, it is somewhat difficult to determine whether our results will generalize to seemingly similar programs that embed ethics education into the CS curriculum, such as the Embedded EthiCS programs at Harvard and Stanford, MIT's SERC program, and Toronto's E3I program, among others. First, our program was developed to serve as a keystone in the larger process of Value Sensitive Design, which likely had the effect of foregrounding values conflicts and the social implications of these cybertechnologies, which in turn may have shifted core theoretical issues in philosophical ethics somewhat to the background, possibly giving them a more interdisciplinary flavor than some other ethics modules taught by philosophers.

Second, largely due to scheduling complications and severe time restrictions caused by the COVID pandemic, some of the more recent VAD modules were constructed and run with only minimal consultation with the CS instructors of record. (Those that were revised versions of earlier modules did benefit second hand from the extensive involvement of CS instructors who assisted with originally setting up the program.) We do believe the program would benefit from having the kind of interdisciplinary team integration that has been achieved at programs like Harvard's, but that kind of close integration mostly proved elusive through the fall of 2021. Instead, our faculty had to rely upon their own expertise related to CS (discussed in Section 3).

Third, whereas most CS ethics module programs utilize a mix of instructors, typically including doctoral candidates and recently minted postdocs, our VAD modules that ran during the semester of this study were taught by two continuing teaching faculty, both of whom had extensive experience teaching related material.

Finally, Northeastern lacks a philosophy graduate student workforce to lean upon for grading, leaving the grading to either CS teaching assistants or the module instructors themselves. Because of this, the learning assessment activities attached to our modules are likely substantially more modest in scope. All of these differences might affect how well our results would generalize, although, in some cases, one might reasonably expect other programs to yield even stronger results (e.g., those able to require additional learning assessment tasks).

Furthermore, all of these limitations only pertain to generalizability to other programs that use embedded modules. It could be that stand alone courses provide even better results, which is an open question.

6.2. Future Directions

We are currently in the process of extending our VAD program in a number of new directions, which we hope will allow us to fill out the lingering questions left by the present study. In particular, through our various partnerships, we are planning to run versions of the modules in CS related courses hosted in new disciplines, at different kinds of institutions, with different levels of instructor experience, in different countries or territories, and possibly also for students at various levels prior to entering college. We are also working to build a program that delivers modules to practitioners in industry, which is a crucial step since most designers and developers of AI, machine learning, and big data tools are already out of university. Testing the modules across these new contexts will be a crucial step in better determining which aspects of our program led to the positive outcomes and whether the results should be expected to transfer.

Finally, we are also exploring whether there are feasible ways to test the effects of VAD modules for the more complex learning goals, like those related to critical ethical reasoning, moral imagination, argument formation, etc. Since we lack the relatively inexpensive workforce associated with a philosophy doctoral program, it is less feasible for us to utilize more traditional methods of assessing progress on these learning goals. That said, there are a number of less traditional tools for assessing ethical reasoning in this space, with varying levels of automation, that have been proposed. These include the Moral Judgment Interview (MJI) (Colby, 1987; Elm and Weber, 1994), the Defining Issues Tests (DIT and DIT-2) (Elm and Weber, 1994; Narvaez and Bock, 2002; Rest, 1975; Rest et al., 1999; Sutton, 1992), the Sociomoral Reflection Measure (SRM) (Basinger et al., 1995; Gibbs et al., 1991) , the Test of Ethical Sensitivity in Science and Engineering (TESSE) (Borenstein et al., 2008), the Engineering and Science Issues Test (ESIT) (Borenstein et al., 2010), and the Engineering Ethical Reasoning Instrument (EERI) (Hess et al., 2019; Zhu et al., 2014). But these tools were all developed for slightly different purposes, some still require extensive amounts of labor to employ, and, to our knowledge, none have undergone the kind of careful philosophical criticism that any such tool ought to be subjected to before widespread use. If a fully automated, feasibly scalable, and philosophically well-grounded assessment method can be developed for assessing progress in the critical preconditions of effective ethical reflection and reasoning, we might have a much better grasp of how effective our modules are across this range of contexts.

Narvaez D and Bock T (2002) Moral Schemas and Tacit Judgement or How the Defining Issues Test is Supported by Cognitive Science. *Journal of Moral Education* 31(3). Routledge: 297–314. DOI: 10.1080/0305724022000008124.

Nissenbaum H (1998) Values in the Design of Computer Systems. *Computers and Society* 28(1): 38–39.

Pitt JC (2014) "Guns Don't Kill, People Kill"; Values in and/or Around Technologies. In: Kroes P and Verbeek P-P (eds) *The Moral Status of Technical Artefacts*. Philosophy of Engineering and Technology. Dordrecht: Springer Netherlands, pp. 89–101. DOI: 10.1007/978-94-007-7914-3_6.

Price PC, Jhangiani R, Chiang I-CA, et al. (2017) One-Group Designs. In: *Research Methods in Psychology*. 3rd ed. Available at: https://opentext.wsu.edu/carriecuttler/chapter/8-1-one-group-designs/ (accessed 4 April 2022).

Quinn MJ (2006) On teaching computer ethics within a computer science department. *Science and Engineering Ethics* 12(2): 335–343. DOI: 10.1007/s11948-006-0032-9.

Raz J (1985) Authority, Law and Morality. *The Monist* 68(3). Oxford University Press: 295–324.

Reis HT and Judd CM (eds) (2014) *Handbook of Research Methods in Social and Personality Psychology*. 2nd edition. New York, NY: Cambridge University Press.

Rest J, Narvaez D, Bebeau M, et al. (1999) A Neo-Kohlbergian Approach: The DIT and Schema Theory. *Educational Psychology Review* 11(4): 291–324. DOI: 10.1023/A:1022053215271.

Rest JR (1975) Longitudinal study of the Defining Issues Test of Moral Judgment: A Strategy for Analyzing Developmental Change. *Developmental Psychology* 11(6). US: American Psychological Association: 738–748. DOI: 10.1037/0012-1649.11.6.738.

Sengers P, Boehner K, David S, et al. (2005) Reflective design. In: *Proceedings of the 4th decennial conference on Critical computing: between sense and sensibility*, 2005, pp. 49–58.

Shilton K (2013) Values Levers: Building Ethics into Design. *Science, Technology, & Human Values* 38(3). SAGE Publications Inc: 374–397. DOI: 10.1177/0162243912436985.

Shilton K (2018) Values and Ethics in Human-Computer Interaction. *Foundations and Trends in Human-Computer Interaction* 12(2): 107–171. DOI: 10.1561/1100000073.

Singer N (2019) The Hard Part of Computer Science? Getting Into Class. *The New York Times*, 24 January. Available at:
23